\documentstyle[11pt,aaspp4]{article}

\def\la{\mathrel{\hbox{\rlap{\hbox{\lower4pt\hbox{$\sim$}}}\hbox{$<$}}}}
\def\ga{\mathrel{\hbox{\rlap{\hbox{\lower4pt\hbox{$\sim$}}}\hbox{$>$}}}}

\slugcomment{Accepted to {\it The Astrophysical Journal}}

\begin{document}

\title{A Possible Cepheid-Like Luminosity Estimator for the Long
Gamma-Ray Bursts}

\author{Daniel E. Reichart\altaffilmark{1,2,3}, Donald Q.
Lamb\altaffilmark{1}, Edward E. Fenimore\altaffilmark{4}, Enrico 
Ramirez-Ruiz\altaffilmark{5}, Thomas L. Cline\altaffilmark{6}, and Kevin 
Hurley\altaffilmark{7}}

\altaffiltext{1}{Department of Astronomy and Astrophysics, University of 
Chicago, 5640 South Ellis Avenue, Chicago, IL 60637}
\altaffiltext{2}{Department of Astronomy, California Institute of Technology, 
Mail Code 105-24, 1201 East California Boulevard, Pasadena, CA 91125}
\altaffiltext{3}{Hubble Fellow}
\altaffiltext{4}{D436, Los Alamos National Laboratory, Los Alamos, New Mexico 
87545} 
\altaffiltext{5}{Institute of Astronomy, University of Cambridge, Cambridge, CB3 
0HA, United Kingdom}
\altaffiltext{6}{NASA Goddard Space Flight Center, Code 661, Greenbelt, MD 
20771}
\altaffiltext{7}{University of California, Berkeley, Space Sciences Laboratory, 
Berkeley, CA 94720}

\begin{abstract}   

We present a possible Cepheid-like luminosity estimator for the long
gamma-ray bursts based on the variability of their light curves.  To
construct the luminosity estimator, we use {\it CGRO}/BATSE data for 13
bursts, {\it Wind}/KONUS data for 5 bursts, {\it Ulysses}/GRB data for
1 burst, and NEAR/XGRS data for 1 burst.  Spectroscopic redshifts, peak
fluxes, and high resolution light curves are available for 11 of these
bursts; partial information is available for the remaining 9 bursts. 
We find that the isotropic-equivalent peak luminosities $L$ of these bursts
positively correlate with a rigorously-constructed measure $V$ of the
variability of their light curves.  We fit a model to these data that
accommodates both intrinsic scatter (statistical variance) and
extrinsic scatter (sample variance).  We find that $L \sim 
V^{3.3^{+1.1}_{-0.9}}$.  If one excludes GRB 980425 from
the fit on the grounds that its association with SN 1998bw at a
redshift of $z = 0.0085$ is not secure, the luminosity estimator spans
$\approx 2.5$ orders of magnitude in $L$, and the slope of the
correlation between $L$ and $V$ is positive with a probability of $1 -
1.4 \times 10^{-4}$ (3.8 $\sigma$).  Although GRB 980425 is excluded from
this fit, its $L$ and $V$ values are consistent with the fitted model,
which suggests that GRB 980425 may well be associated with SN 1998bw,
and that GRB 980425 and the cosmological bursts may share a common
physical origin.  If one includes GRB 980425 in the fit, the luminosity
estimator spans $\approx 6.3$ orders of magnitude in $L$, and the slope
of the correlation is positive with a probability of $1 - 9.3 \times
10^{-7}$ (4.9 $\sigma$).  In either case, the luminosity estimator yields
best-estimate luminosities that are accurate to a factor of $\approx
4$, or best-estimate luminosity distances that are accurate to a factor
of $\approx 2$.  Independently of whether or not GRB 980425 should be
included in the fit, its light curve is unique in that it is much less
variable than the other $\approx 17$ light curves of bursts in our
sample for which the signal-to-noise is reasonably good.

\end{abstract}


\section{Introduction}

Since gamma-ray bursts (GRBs) were first discovered (Klebesadel,
Strong, \& Olson 1973), thousands of bursts have been detected by a
wide variety of instruments, most notably, the Burst and Transient
Source Experiment (BATSE) on the {\it Compton Gamma-Ray Observatory
(CGRO)}, which will have detected about 2700 bursts by the end of {\it
CGRO}'s more than 9 year mission in 2000 June (see, e.g., Paciesas et
al. 1999).  However, the distance scale of the bursts remained
uncertain until 1997, when BeppoSAX began localizing long bursts to a
few arcminutes on the sky, and distributing the locations to
observers within hours of the bursts.  This led to the discovery of
X-ray (Costa et al. 1997), optical (van Paradijs et al. 1997), and
radio (Frail et al. 1997) afterglows, as well as host galaxies (Sahu et
al. 1997).  Subsequent observations led to the spectroscopic
determination of burst redshifts, using absorption lines in the spectra
of the afterglows (see, e.g., Metzger et al. 1997), and emission lines
in the spectra of the host galaxies (see, e.g., Kulkarni et al. 1998a). 
To date, redshifts have been measured for 13 bursts.

Recently, Stern, Poutanen, \& Svensson (1999; see also Stern, Svensson,
\& Poutanen 1997), Norris, Marani \& Bonnell (2000; see also Norris,
Marani \& Bonnell 1999), and Fenimore \& Ramirez-Ruiz (2000; see also
Ramirez-Ruiz \& Fenimore 1999) have proposed trends between burst
luminosity and quantities that can be measured directly from burst
light curves, for the long bursts.  Using 1310 BATSE bursts for which
peak fluxes and high resolution light curves were available, Stern,
Poutanen, \& Svensson (1999) have suggested that simple bursts (bursts
dominated by a single, smooth pulse) are less luminous than complex
bursts (bursts consisting of overlapping pulses); however, see \S 5.  Using a 
sample of 7
BATSE bursts for which spectroscopic redshifts, peak fluxes, and high
resolution light curves were available, Norris, Marani \& Bonnell
(2000) have suggested that more luminous bursts have shorter spectral
lags (the interval of time between the peak of the light curve in
different energy bands).  Using the same 7 bursts, Fenimore \&
Ramirez-Ruiz (2000) have suggested that more luminous bursts have more
variable light curves.  These trends between luminosity and quantities that can 
be measured directly from light curves raise the exciting possibility that 
luminosities, and hence luminosity distances, might be inferred for the long 
bursts from their light curves alone.

In this paper, we present a possible luminosity estimator for the long
bursts, the construction of which was motivated by the work of
Ramirez-Ruiz \& Fenimore (1999) and Fenimore \& Ramirez-Ruiz (2000). 
We term the luminosity estimator ``Cepheid-like'' in that it can be
used to infer luminosities and luminosity distances for the long bursts
from the variabilities of their light curves alone.  We apply this
luminosity estimator to every long burst in the current BATSE
catalog in an upcoming paper (Reichart et al. 2000).  

We rigorously construct our measure $V$ of the variability of a burst
light curve in \S 2.  Qualitatively, $V$ is computed by taking the
difference of the light curve and a smoothed version of the light
curve, squaring this difference, summing the squared difference over
time intervals, and appropriately normalizing the result.  Our
variability measure differs from that of Fenimore \& Ramirez-Ruiz
(2000) in three important ways: (1) we define the timescale over which
the light curve is smoothed differently; (2) we subtract out the
contribution to the variability due to Poisson noise differently; and
(3) we combine variability measurements of light curves acquired in
different energy bands into a single measurement of a burst's
variability differently.  

We find that only smoothing timescales that are proportional to burst
duration appear to lead to significant correlations between the 
isotropic-equivalent peak photon luminosity $L$ of a burst and
$V$; in particular, smoothing timescales that are a fixed duration in
the source frame do not.  We take the smoothing timescale of a burst
light curve, acquired in observer-frame energy band $E$, to be the time
spanned by the brightest $100f$\% of the total counts above background,
where $f$ is a number between 0 and 1 that we fix using the data in \S
4.  We schematically illustrate this measure of duration in Figures 1
and 2.  In Figure 1, the 50\% smoothing timescale of the light curve is
given by $T_{f=0.5}^E = T_1 + T_2$, and the 90\% smoothing timescale of
the light curve is given by $T_{f=0.9}^E = T_3$; in Figure 2, the 50\%
smoothing timescale of the light curve is given by $T_{f=0.5}^E = T_4$,
and the 90\% smoothing timescale of the light curve is given by
$T_{f=0.9}^E = T_5 + T_6$.  We have chosen this measure of the
smoothing timescale over others because it is robust:  small variations
in either the light curve or the value of $f$ result in only small
variations in the value of the smoothing timescale.  This is not the
case with measures like $T_{50}$ and $T_{90}$ (see, e.g., Paciesas et
al. 1999).  

For example, consider the case of a burst with a
precursor.  The value of, say, $T_{90}$ can differ considerable if the
precursor's counts above background is $< 5$\% versus $> 5$\% of the
total counts above background.  Likewise, if the precursor's counts
above background is, say, 5\% of the total counts above background, the
duration that one measures can differ considerably if one uses
$T_{<90}$ versus $T_{>90}$.  The effect of using a smoothing timescale that is 
artificially long (short) is measuring a variability that is artificially high 
(low).  Our measure of the smoothing timescale does not suffer from such 
robustness problems.

We present our measure of the isotropic-equivalent peak photon luminosity $L$ of 
a
burst in \S 3.  In \S 4, we expand the original Ramirez-Ruiz \&
Fenimore (1999) sample of 7 bursts to include a total of 20 bursts,
including 13 BATSE bursts, 5 {\it Wind}/KONUS bursts, 1 {\it
Ulysses}/GRB burst, and 1 NEAR/XGRS burst.  Spectroscopic redshifts,
peak fluxes, and high resolution light curves are available for 11 of
these bursts; partial information is available for the remaining 9
bursts.  

Also in \S 4, we construct our luminosity estimator, which differs from
that of Fenimore \& Ramirez-Ruiz (2000) in two important ways:  (1)
applying the Bayesian inference formalism developed by Reichart (2000),
we fit a model to the data that accommodates both intrinsic scatter
(statistical variance) in two dimensions and extrinsic scatter (sample
variance) in two dimensions; and (2) again applying this Bayesian
inference formalism, we determine not only
the best estimate for $L$ as a function of $V$, but also the
uncertainty in $L$ as a function of $V$, as well as approximate these
functions with analytic expressions.  We state our conclusions in \S 5.

\section{The Variability Measure}

We now rigorously construct a measure of the variability of a burst
light curve.  We require it to have the following properties:  (1) we
define it in terms of physical, source-frame quantities, as
opposed to measured, observer-frame quantities; (2) when converted to
observer-frame quantities, all strong dependences on redshift and other
difficult or impossible to measure quantities cancel out; (3) it is not
biased by instrumental binning of the light curve, despite cosmological
time dilation and the narrowing of the light curve's temporal
substructure at higher energies (Fenimore et al. 1995); (4) it is not
biased by Poisson noise, and consequently can be applied to faint bursts; and 
(5) it is robust; i.e., similar light curves always
yield similar variabilities.  We also derive an expression for the
statistical uncertainty in a light curve's measured variability. 
Finally, we describe how we combine variability measurements of light
curves acquired in different energy bands into a single measurement of
a burst's variability. 

We first define the variability of a burst light curve, acquired in
observer-frame energy band $E$, in terms of physical, source-frame
quantities: 
\begin{equation}
V_f^E = \frac{\int_{-\infty}^{\infty}[L^{E_s}(t_s)-(L^{E_s}\star
S_f)(t_s)]^2dt_s}{\int_{-\infty}^{\infty} [L^{E_s}(t_s)]^2dt_s},
\label{physvar}
\end{equation}
where $L^{E_s}(t_s)$ is the luminosity of the burst in source-frame
energy band $E_s = E(1+z)$ as a function of source-frame time $t_s$,
and $(L^{E_s}\star S_f)(t_s)$ is the convolution of this function and a
boxcar smoothing function of area equal to 1 and width equal to the
smoothing timescale of $L^{E_s}(t_s)$, i.e., the source-frame smoothing
timescale of the light curve.  Since our definition of the smoothing
timescale is a robust measure of duration (\S 1), Equation
(\ref{physvar}) is a robust measure of variability.

Next, we convert Equation (\ref{physvar}) to observer-frame quantities:
\begin{equation}
V_{f,P}^E = \frac{\int_{-\infty}^{\infty} \{C^E(t_o) - B^E(t_o) - [(C^E - 
B^E)\star S_f](t_o)\}^2[\Delta \Omega 
D^2(z)(1+z)^{\alpha}]^2(1+z)^{\beta}dt_o}{\int_{-\infty}^{\infty} [C^E(t_o) - 
B^E(t_o)]^2[\Delta \Omega D^2(z)(1+z)^{\alpha}]^2(1+z)^{\beta}dt_o}.
\label{measvar}
\end{equation}
Here, $C^E(t_o)$ and $B^E(t_o)$ are the total (source plus background)
and background photon count rates in observer-frame energy band $E$ as
a function of observer-frame time $t_o$, $[(C^E-B^E)\star S_f](t_o)$ is
the convolution of $C^E(t_o)-B^E(t_o)$ and a boxcar smoothing function
of area equal to 1 and width equal to the smoothing timescale of
$C^E(t_o)-B^E(t_o)$, i.e., $T_f^E$, $\Delta \Omega$ is the effective
solid angle of the emitted light, $D(z)$ is the comoving distance to
the burst at redshift $z$, $\alpha$ is either 1 or 2 depending on
whether $L^{E_s}(t_s)$ is a photon number or energy luminosity, and
$\beta \approx 1-0.4=0.6$:  the factor of $(1+z)^1$ is due to
cosmological time dilation, and the factor of $\approx (1+z)^{-0.4}$ is
due to the narrowing of the light curve's temporal substructure at
higher energies (Fenimore et al. 1995).  By construction, the strong
dependences on $\Delta \Omega$, $z$, $\alpha$, and $\beta$, and a weak
dependence on the cosmological model, cancel
out.  Furthermore, since $B^E(t_o)$ always varies over much longer
timescales than $T_f^E$, Equation (\ref{measvar}) simplifies to:
\begin{equation}
V_{f,P}^E = \frac{\int_{-\infty}^{\infty} [C^E(t_o) - (C^E\star 
S_f)(t_o)]^2dt_o}{\int_{-\infty}^{\infty} [C^E(t_o) - B^E(t_o)]^2dt_o}.
\label{contvar}
\end{equation}
We distinguish Equations (\ref{measvar}) and (\ref{contvar}) from
Equation (\ref{physvar}) with a subscript $P$ because they differ in
one fundamental way:  unlike $L^{E_s}(t_s)$ in Equation
(\ref{physvar}), $C^E(t_o)$ suffers from Poisson noise.  This
additional source of variability biases Equations (\ref{measvar}) and
(\ref{contvar}) to higher values.  Although this bias is minor for
bursts that are significantly brighter than the background, it cannot
be ignored for faint bursts.  We determine and subtract out the
contribution to the variability due to Poisson noise below.

However, before we can subtract out the contribution to the variability
due to Poisson noise, we must first bin the light curve, since observed
light curves typically are discrete, not continuous.  Consequently, we replace
the integrals in Equation (\ref{contvar}) with sums from $i = 1$ to
$N$, $C^E(t_o)$ with $C_i^E$, $B^E(t_o)$ with $B_i^E$, and $(C^E\star
S_f)(t_o)$ with its discrete equivalent, $S_i(C_j^E,N_f^E)$.  Here,
$C_j^E$ is the light curve to be smoothed, $N_f^E$ is the smoothing
timescale, i.e., $T_f^E$, but measured in number of bins, and 
\begin{equation}
S_i(C_j^E,N_x) = 
\frac{1}{N_x}\left[\sum_{j=i-n_x}^{i+n_x}C_j^E+\left(\frac{N_x-1}{2}-n_x\right)(
C_{i-n_x-1}^E+C_{i+n_x+1}^E)\right],
\label{smear}
\end{equation}
where $n_x$ is the truncated integer value of $(N_x-1)/2$.  Binning can
bias the variability measure, since binning wipes out all variability
on timescales shorter than the sampling timescale of the light curve,
and the effective sampling timescale of the light curve in the source
frame decreases with increasing burst redshift:  although the sampling
timescale of the light curve is fixed by the instrument in the observer
frame, (1) cosmological time dilation and (2) the narrowing of the
light curve's temporal substructure at higher energies (Fenimore et al.
1995) decrease the effective sampling timescale of the light curve in
the source frame.  We remove this redshift bias by smoothing all light
curves on a source-frame timescale of 64 ms (the shortest sampling
timescale of BATSE), which corresponds to a timescale of
$64(1+z)^{\beta}$ ms in the observer frame.  Consequently, we replace
$C_i^E$ with $S_i(C_j^E,N_z)$, where $N_z = 64(1+z)^{\beta}/{\Delta
t}$, and $\Delta t$ is the observer-frame sampling timescale of the
light curve in milliseconds.  Equation (\ref{contvar}) becomes:
\begin{equation}
V_{f,P}^E = \frac{\sum_{i=1}^N [S_i(C_j^E,N_z) - 
S_i(C_j^E,N_f^E)]^2}{\sum_{i=1}^N [S_i(C_j^E,N_z) - B_i^E]^2},
\label{discvar}
\end{equation}
where we take $C_j^E$ and $B_i^E$ to be measured in counts, not counts
per unit time.   Although the computation of $V_{f,P}^E$ now depends on
$z$ and $\beta$, these dependences are very weak:  wide variations in
the values of these parameters do not significantly change the measured
variabilities of the bursts in our sample; i.e., burst light curves
have very little power on such short timescales.  Likewise, wide
variations in the value of the effective sampling timescale of the
light curve that we impose in the source frame also have little effect
on the measured variabilities of these bursts. 

Now, we determine and subtract out the contribution of Poisson noise to
the variability, and simultaneously we derive an expression for the
statistical uncertainty in a light curve's measured variability. 
First, we rewrite the expressions $S_i(C_j^E,N_z) - S_i(C_j^E,N_f^E)$
and $S_i(C_j^E,N_z) - B_i^E$ from Equation (\ref{discvar}) as weighted
sums of the statistically independent measurements $C_j$:
\begin{equation}
V_{f,P}^E = \frac{\sum_{i=1}^N (\sum_{j=1}^Na_{ij}C_j)^2}{\sum_{i=1}^N 
(\sum_{j=1}^Nb_{ij}C_j-B_i)^2},
\label{discvar2}
\end{equation}
where $a_{ij}$ and $b_{ij}$ are weights that differ for each burst, but
can be computed straightforwardly using Equation (\ref{smear}).  Since
we are in the Gaussian limit, the uncertainty in each $C_j$ is simply
$\pm\sqrt{C_j}$.  Since each $C_j$ is statistically independent, their
weighted uncertainties can be summed in quadrature.  Hence, the sums
over $j$ in Equation (\ref{discvar2}) and their uncertainties are:  
$\sum_{j=1}^Na_{ij}C_j\pm\sqrt{\sum_{j=1}^Na_{ij}^2C_j}$ and
$\sum_{j=1}^Nb_{ij}C_j-B_i\pm\sqrt{\sum_{j=1}^Nb_{ij}^2C_j}$.  

Next, in accordance with Equation (\ref{discvar2}), we square these
expressions, which yields expressions consisting of three terms:  the
square of the original term, a positive term due to Poisson noise, and
an uncertainty term.  Subtracting out the contributions to Equation
(\ref{discvar2}) due to Poisson noise yields:
\begin{equation}
V_f^E = \frac{\sum_{i=1}^N 
[(\sum_{j=1}^Na_{ij}C_j)^2-\sum_{j=1}^Na_{ij}^2C_j]}{\sum_{i=1}^N 
[(\sum_{j=1}^Nb_{ij}C_j-B_i)^2-\sum_{j=1}^Nb_{ij}^2C_j]}, 
\label{var}
\end{equation}
where the uncertainty in the $i$th term of the numerator is
$\pm2(\sum_{j=1}^Na_{ij}C_j)\sqrt{\sum_{j=1}^Na_{ij}^2C_j}$, and the
uncertainty in the $i$th term of the denominator is 
$\pm2(\sum_{j=1}^Nb_{ij}C_j-B_i)\sqrt{\sum_{j=1}^Nb_{ij}^2C_j}$.
Recognizing that only $\approx N/N_z$ of the terms in the sums over $i$
are statistically independent, since the sums over $j$ correspond to a
convolution of the light curve and a boxcar smoothing function of width
equal to $N_z$ bins, the uncertainties in the sums over $i$ are a
factor of $\approx \sqrt{N_z}$ larger than what would be derived if all
of the terms were statistically independent.  Hence, the uncertainty in
the numerator is $\approx
\pm2\sqrt{N_z\sum_{i=1}^N(\sum_{j=1}^Na_{ij}C_j)^2\sum_{j=1}^Na_{ij}^2C_j}$,
and the uncertainty in the denominator is $\approx
\pm2\sqrt{N_z\sum_{i=1}^N(\sum_{j=1}^Nb_{ij}C_j-B_i)^2\sum_{j=1}^Nb_{ij}^2C_j}$.
These uncertainties can be at most only weakly correlated.  Taking them to be
independent, we find that the statistical uncertainty in a light
curve's measured variability is approximately given by:
\begin{equation}
\sigma_{V_f^E} = 
V_f^E\sqrt{\frac{4N_z\sum_{i=1}^N(\sum_{j=1}^Na_{ij}C_j)^2\sum_{j=1}^Na_{ij}^2C_
j}{\{\sum_{i=1}^N 
[(\sum_{j=1}^Na_{ij}C_j)^2-\sum_{j=1}^Na_{ij}^2C_j]\}^2}+\frac{4N_z\sum_{i=1}^N(
\sum_{j=1}^Nb_{ij}C_j-B_i)^2\sum_{j=1}^Nb_{ij}^2C_j}{\{\sum_{i=1}^N 
[(\sum_{j=1}^Nb_{ij}C_j-B_i)^2-\sum_{j=1}^Nb_{ij}^2C_j]\}^2}}.
\label{intscat}
\end{equation}

All that remains is to describe how we combine variability measurements
of light curves acquired in different energy bands, $V_f^E$, into a
single measurement of a burst's variability:  $V_f$.  For each burst in
our sample, light curves were acquired in typically 3 or 4
independent energy bands (see Table 2).  We find that (1) the smoothing 
timescales of
these bursts' light curves decrease with energy as $\approx E^{-0.4}$
(Figure 3), and (2) the variabilities of these bursts' light curves are
approximately constant across energy bands (Figure 4).  In hindsight,
the former result is not surprising, given our definition of the
smoothing timescale and the principle result of Fenimore et al. (1995);
however, this does constitute an independent confirmation of their
result.  The latter result suggests that we can model a burst's
variability as a constant across energy bands.  Applying the Bayesian
inference formalism developed by Reichart (2000) for fitting models
with extrinsic scatter (sample variance) to data with intrinsic scatter
(statistical variance, in this case given by Equation \ref{intscat}),
we fit this model to the typically 3 or 4 independent
measurements of a burst's variability made in independent energy bands,
$V_f^E$, resulting in a single measurement of that burst's variability,
$V_f$, and the uncertainty in $V_f$.  We plot the $> 25$ keV light curves of the 
most and least variable cosmological BATSE bursts in our sample in Figure 5.

\section{The Luminosity Measure}

Let $P$ be the peak flux of a burst in photons cm$^{-2}$ s$^{-1}$
between observer-frame energies $E_l$ and $E_u$.  The
isotropic-equivalent peak photon luminosity of the burst in erg s$^{-1}$
between source-frame energies 100 and 1000 keV is given by 
\begin{equation} L = 4\pi D^2(z) P
\frac{\int_{100}^{1000}E\Phi\left(\frac{E}{1+z}\right)dE}{\int_{E_l}^{E_u}\Phi(E
)dE},
\end{equation}  where $\Phi(E)$ is the observer-frame spectral shape,
which we parameterize with the Band function (Band et al. 1993).  For
each burst in our sample, we compute the value of $L$ for each of the
54 parameterizations of the Band function in Band et al. (1993); the
luminosity error bars in Figures 9 and 10 are dominated by this
variation in the parameterization of the Band function, but clearly are
too small to matter.  Hence, $L$ is very insensitive to reasonable
variations in the parameterization of the Band function for the bursts
in our sample.  The choice of source-frame energy range also is very
unimportant:  we chose 100 -- 1000 keV to approximately match the
observer-frame energy range in which BATSE measures peak fluxes, 50 --
300 keV, for redshifts that are typical of the bursts in our sample: 
$z \approx 1 - 2$.

\section{The Luminosity Estimator}

We list our sample of 20 bursts in Table 1; it consists of every
burst for which redshift information is currently available.  Peak
fluxes are available for all 20 bursts.  Spectroscopic redshifts and
64-ms or better resolution light curves are available for 11 bursts. 
Spectroscopic redshifts but only 1-s resolution light curves are
available for 2 bursts.  We compute variability lower limits for these
bursts:  we compute variabilities from their 1-s resolution light
curves without further degrading the effective resolution of these
light curves by smoothing them on the 64-ms source-frame timescale (\S
2); these variabilities are lower limits to the variabilities that we
would compute if 64-ms or better resolution light curves were
available.  Redshift upper limits (1) from the non-detection of the
Ly$\alpha$ forest in host galaxy spectroscopy in the case of 1 burst,
and (2) from the non-detection of the Lyman limit in afterglow and host
galaxy photometry in the case of 6 bursts, and 64-ms or better
resolution light curves are available for the remaining 7 bursts.  We
compute luminosity upper limits for these bursts.  We compute
variabilities for these bursts for both $z = 0$ and $z$ equal to the
redshift upper limit; for all 7 bursts, both of these variabilities are
nearly identical, testifying to the weakness of the computational
dependence of the variability on redshift (\S 2).  In addition to the
data listed in Table 1, we used {\it Ulysses}/GRB data for GRBs 970228,
990712, and 991208 to test the consistency of our results across
instruments; we find the measured variabilities and luminosities of
these bursts to be fully consistent across instruments. 

We now construct the luminosity estimator.  First, we compute the
variabilities, $V_f$, and the isotropic-equivalent peak luminosities,
$L$,\footnote{We use peak fluxes measured on a 1-s timescale, and we
take $H_0 = 65$ km s$^{-1}$ Mpc$^{-1}$, $\Omega_m = 0.3$, and
$\Omega_{\Lambda} = 0.7$; the luminosity estimator is very
insensitive to these choices.} of the bursts in our sample.  We compute
the variabilities as functions of $f$, where we use the value of $f$ to
compute the smoothing timescales, $T_f^E$, of all of the light curves
in our sample (\S 1).  We identify values of $f$ that lead to
successful and robust luminosity estimators below.  Regardless of the
value of $f$ that we use to compute the variabilities, the distribution
of our sample's bursts in the $\log{L}$-$\log{V_f}$ plane appears to be
well modeled by a normal distribution about a straight line (see, e.g.,
Figure 9).  We parameterize the line by:
\begin{equation}   
\log{V_f}(L) = \log{{\bar{V}}_f} + b + m(\log{L} - \log{\bar{L}}), 
\label{line}
\end{equation}   
where $b$ is the intercept of the line, $m$ is its slope, and
${\bar{V}}_f$ and $\bar{L}$ are the median values of $V_f$ and $L$ for
the bursts in our sample for which spectroscopic redshifts, peak
fluxes, and 64-ms or better resolution light curves are available.  We
parameterize this line as a function of $L$, instead of as a function
of $V_f$, because the data do not fully rule out the possibility of $m
\la 0$, whereas the data do rule out the possibility of very large
values of $m$.  We parameterize the normal distribution about this line
by $\sigma_{\log{V_f}}$, which is half of the distribution's 1-$\sigma$ width 
along the $\log{V_f}$ axis.  Applying the Bayesian inference formalism
developed by Reichart (2000) for fitting data with extrinsic scatter
(in this case, $\sigma_{\log{V_f}}$) to data with intrinsic scatter (in
this case, the uncertainties in the measured values of $V_f$ and $L$),
we determine values and uncertainties for the model parameters ($b$,
$m$, and $\sigma_{\log{V_f}}$) as functions of $f$ (Figure
6).\footnote{This Bayesian inference formalism deals only with measurements with 
Gaussian error distributions,
not with lower or upper limits.  However, this formalism can be 
straightforwardly generalized to deal with limits as well, using two
facts:  (1) a limit can be given by the convolution of a Gaussian
distribution and a Heavyside function, and (2) convolution is
associative.  In this paper, we fit to both measurements and limits.}

Of the 20 bursts in our sample, GRB 980425 is unique due to its
possible association with SN 1998bw (Kulkarni et al. 1998b; Galama et
al. 1998) at a redshift of $z = 0.0085$ (Tinney et al. 1998); the 12
other bursts for which spectroscopic redshifts are available have
$0.430 < z < 3.418$.  Consequently, we first construct the luminosity
estimator excluding GRB 980425 from the above fits; we then repeat the
fits including this burst.  Excluding GRB 980425 from the fits, we
first identify values of $f$ that lead to successful and robust
luminosity estimators.  One measure of the success of a luminosity
estimator is the probability that its slope, $m$, departs from $m =
0$.  Since the best-fit values of $m$ are positive (Figure 6), we
compute and plot in Figure 7 the probability that $m < 0$, $P(m<0)$, as
a function of $f$.   We find that values of $f$ between $\approx 0.2$
and $\approx 0.5$ lead to luminosity estimators with slopes that are
positive with approximately equally probability:  $P(m<0) \approx 1.4
\times 10^{-4}$ (3.8 $\sigma$).  Values of $f \la 0.2$ and $f \ga 0.5$
lead to luminosity estimators with slopes that are positive with
increasingly less probability.  In addition to having a slope that is
positive (or negative) with large probability, a successful luminosity
estimator also should be robust:  small changes in the value of $f$
should lead to only small changes in the estimated values of $L$.  
Consequently, we also plot in Figure 7 the extrinsic scatter of the
data along the $\log{L}$ axis, $\sigma_{\log{L}} =
\sigma_{\log{V_f}}/m$, as a function of $f$; this function is a measure of the 
maximum amount that the estimated luminosities of the bursts in our
sample can change as a function of $f$.  We find that values of $f \ga
0.5$ lead to increasingly less robust luminosity estimators.  
Taking $f = 0.45$, which corresponds to the minimum in
$P(m<0)$, we compute smoothing timescales, $T_f^{E_i}$, for each of the 
instrument's available energy bands, $E_i$, (Table 2), and using these smoothing 
timescales, we find that that $\log{{\bar{V}}_f} = -1.088$, $\log{\bar{L}} =
51.852$, $b = 0.013_{-0.092}^{+0.075}$, $m = 0.302_{-0.075}^{+0.112}$,
and $\sigma_{\log{V_f}} = 0.175_{-0.046}^{+0.073}$.  Hence, we find that $L \sim 
V_{f=0.45}^{3.3^{+1.1}_{-0.9}}$.  We plot the
two-dimensional uncertainty distributions of the fitted values of the
model parameters in Figure 8.  We plot the data and the best-fit model
of the distribution of these data in the $\log{L}$-$\log{V_f}$ plane
for $f = 0.45$ in Figure 9.

Although the solid line in Figure 9 is a good approximation to the best
estimate for $L$ as a function of $V_f$, the dotted lines in Figure 9
do not correspond to the uncertainty in $L$ as a function of $V_f$.  The dotted 
lines in Figure 9 mark the best-fit width, given by $\sigma_{\log{V_f}}
= 0.175$, of the best-fit model of the distribution of the data in the
$\log{L}$-$\log{V_f}$ plane; they do not account for the uncertainties
in the fitted values of the model parameters, and the uncertainties in
the fitted values of $b$ and $m$ in particular.  Applying the Bayesian
inference formalism of Reichart (2000), we formally compute the best
estimate for $L$ as a function of $V_f$, and the uncertainty in $L$ as
a function of $V_f$, which we plot as the solid curves in Figure 10. 
The computation of this distribution in the $\log{L}$-$\log{V_f}$ plane
is numerically intensive, so we also provide analytic approximations to
these functions, which we plot as the dotted curves in Figure 10.  The
best estimate for $L$ as a function of $V_f$ can be approximated by
Equation (\ref{line}), using the best-fit values of the model
parameters.  Approximating the uncertainty distributions of the fitted
values of the model parameters (Figure 8) as uncorrelated and Gaussian,
the uncertainty in $L$ as a function of $V_f$ can be approximated by
(Reichart 2000):
\begin{equation}
\sigma_{\log{V_f}}(L) = \sqrt{\sigma_{\log{V_f}}^2 + \sigma_b^2 + 
\sigma_m^2(\log{L}-\log{\bar{L}})^2},
\label{lumerr}
\end{equation}
where $\sigma_{\log{V_f}}(L)$ is the 1-$\sigma$ uncertainty in $V_f$ as
a function of $L$, $\sigma_{\log{V_f}} = 0.175$, $\sigma_b \approx
0.084$ is the uncertainty in the fitted value of $b$, and $\sigma_m
\approx 0.094$ is the uncertainty in the fitted value of $m$.  At its
best, the luminosity estimator constructed excluding GRB 980425 from
the fits yields best-estimate luminosities that are accurate to
a factor of $\approx 4$, or best-estimate luminosity
distances that are accurate to a factor of $\approx 2$.

In addition to being used to estimate the luminosities of bursts for
which spectroscopic redshifts are not available, and the uncertainties
in these luminosities, Figure 10 also can be used to determine if other
bursts for which spectroscopic redshifts are available, such as GRB
980425 or future bursts, are consistent with the above-fitted model. 
Consequently, we also plot GRB 980425 in Figure 10.  Although its
implied luminosity, $L = 5.2\pm2.0\times10^{46}$ erg s$^{-1}$, is
$\approx 4 - 6$ orders of magnitude less than the 12 other measured luminosities 
of bursts in our sample, its variability, $V_{f = 0.45} =
0.00\pm0.01$, also is considerably less than the $\approx 17$ other reasonably 
well-constrained variabilities of bursts in our sample, and clearly is
consistent with the above-fitted model.  We plot the $> 25$ keV BATSE light 
curve of GRB 980425 in Figure 11.  Repeating the fits including
GRB 980425 yields:  $f = 0.47$, $\log{{\bar{V}}_f} = -1.073$, $\bar{L}
= 51.697$, $b = -0.035_{-0.089}^{+0.080}$, $m =
0.306_{-0.071}^{+0.099}$, $\sigma_{\log{V_f}} =
0.170_{-0.041}^{+0.074}$, and $P(m<0) = 9.3 \times
10^{-7}$ (4.9 $\sigma$) (Figures 6 and 7).  Hence, we find that $L \sim 
V_{f=0.47}^{3.3^{+1.0}_{-0.8}}$.  We also plot the two-dimensional uncertainty
distributions of the fitted values of the model parameters in Figure
8.  We plot the data and the best-fit model of the distribution of
these data in the $\log{L}$-$\log{V_f}$ plane for $f = 0.47$ in Figure
12.  We plot the best estimate for $L$ as a function of $V_f$, the
uncertainty in $L$ as a function of $V_f$, and our analytic
approximations to these functions, given by Equations (\ref{line}) and
(\ref{lumerr}) and the fitted values of the model parameters, in Figure 13.

\section{Discussion and Conclusions}

We have presented a rigorously-constructed measure of the variability
of a burst's light curve.  Using this variability measure and a sample
of 20 bursts, consisting of every burst for which redshift information
is currently available, we have shown that a significant correlation
exists between the variability of a burst's light curve, and the
burst's isotropic-equivalent peak photon luminosity.  This correlation
between variability and luminosity is in agreement with the trends
found by Stern, Poutanen, \& Svensson (1999) and Fenimore \&
Ramirez-Ruiz (1999).  That is, more variable (``complex'' in the
terminology of Stern, Poutanen, \& Svensson 1999) bursts are more
luminous, while less variable (``simple'' in the terminology of Stern,
Poutanen, \& Svensson 1999) bursts are less luminous.\footnote{However, we must 
draw attention to a potential disagreement between the primary
conclusion of Stern, Poutanen, \& Svensson (1999), and the width of the
luminosity distribution of the multiply-peaked bursts with
spectroscopically-measured redshifts in our sample.  Stern, Poutanen, \&
Svensson (1999) find that the differential peak count rate distribution of their
complex, or multiply-peaked BATSE bursts peaks about a factor of 4 above
threshold, while the differential peak count rate distribution of their simple,
or singly-peaked BATSE bursts does not have a similar peak.  They interpret this
to mean that their complex bursts are more luminous and at higher redshifts than
their simple bursts:  they argue that this peak corresponds to the peak in the
star-formation history of the universe at $z \sim 1.5$, and that it is not a
threshold effect.  However, as this peak is narrow, spanning less than an order
of magnitude, this could only be the case if the luminosity function of the
complex bursts is similarly narrow; otherwise, this feature would be washed out.
We have visually examined the light curves of the bursts with
spectroscopically-measured redshifts in our sample, and find that only GRB
980425 and GRB 970508 appear to be singly-peaked.  The remaining,
multiply-peaked bursts span $\approx 2.5$ orders of magnitude, which appears to 
contradict their interpretation of this peak that they find in the differential 
peak count rate distribution of their complex bursts.}

Furthermore, from the correlated variabilities and luminosities of our
sample of 20 bursts, we have constructed a possible Cepheid-like
luminosity estimator for the long bursts.  If one excludes GRB 980425
from the fits, the luminosity estimator spans $\approx 2.5$ orders of
magnitude in luminosity, and its slope is positive with a probability
of $1 - 1.4 \times 10^{-4}$ (3.8 $\sigma$).  GRB 980425, however, is
consistent with the fitted model.  If one includes this burst in the
fits, the luminosity estimator spans $\approx 6.3$ orders of magnitude
in luminosity, and its slope is positive with a probability of $1 - 9.3
\times 10^{-7}$ (4.9 $\sigma$). 

Future bursts will either increase these probabilities, or
possibly disprove the luminosity estimator.  However, since the
uncertainty in $L$ as a function of $V_f$ in Figures 10 and 13 is
dominated by extrinsic scatter (i.e., $\sigma_{\log{V_f}}$), and not by
the uncertainties in the fitted values of the model parameters, a
larger sample of bursts is unlikely to improve the predictive power of
the luminosity estimator.  Currently, the luminosity estimator yields
best-estimate luminosities that are accurate to a factor of $\approx 4$, or 
best-estimate luminosity distances that are accurate to a
factor of $\approx 2$.

However, independently of whether or not the luminosity estimator is eventually 
disproved, the light curve of GRB 980425 is unique
in that it is much less variable than the other $\approx 17$ light
curves of bursts in our sample for which the signal-to-noise is reasonably good: 
 $\log{V_{f=0.45}} <
-1.5$ with a probability of $1 - 3.4 \times 10^{-4}$ (3.4 $\sigma$), and
$\log{V_{f=0.45}} < -1$ with a probability of $1 - 3.5 \times 10^{-23}$ (9.9 
$\sigma$).  The argument has been made that the association of GRB
980425 with SN 1998bw at the unusually low redshift of $z = 0.0085$ 
probably is accidental because the light curve of GRB 980425 is no
different than the light curves of the cosmological bursts.  On the
contrary, we find that the light curve of GRB 980425 is different from the
light curves of the cosmological bursts.  Consequently, GRB 980425 may well be 
associated with SN 1998bw.

If GRB 980425 is associated with SN 1998bw, and if the luminosity
estimator is correct, the fact that GRB 980425 is consistent with the
fitted model suggests that
GRB 980425 and the cosmological bursts may share a common, or at least
a related, physical origin, although they cannot share a common redshift
distribution and/or luminosity function (Graziani, Lamb, \& Marion 1999).  This
conclusion is made more intriguing by the recent discoveries of
supernova-like components to the late afterglows of the cosmological
bursts GRB 970228 (Reichart 1999, Galama et al. 2000; Reichart,
Castander, \& Lamb 2000; Reichart, Lamb, \& Castander 2000) and GRB
980326 (Bloom et al. 1999).

\acknowledgements
Support for this work was provided by NASA through the Hubble Fellowship grant
\#HST-SF-01133.01-A from the Space Telescope Science Institute, which is
operated by the Association of Universities for Research in Astronomy, Inc.,
under NASA contract NAS5-26555.  Support for this work was also provided by NASA 
contracts NASW-4690 and SCSV 464006.  K. H. acknowledges {\it Ulysses} support 
under JPL Contract 958056.  We are very grateful to Evgeny Mazets and the KONUS 
team for granting us the use of their data.  We are also grateful to Paul 
Butterworth for retrieving the KONUS data, and Marc Kippen for retrieving the 
BATSE data for GRB 980613.  We also thank Carlo Graziani for generously sharing 
his insights into Bayesian inference.  Finally, D. E. R. thanks Bob Nichol and 
the Department of Physics at Carnegie Mellon University, and Dr. and Mrs. 
Keisler for their hospitality during the summer of 2000.

\clearpage

\begin{deluxetable}{cccccc}
\footnotesize
\tablecolumns{5}
\tablewidth{0pc}
\tablecaption{Redshifts, Luminosities, and Variabilities}
\tablehead{\colhead{GRB} & \colhead{$z$} &
\colhead{$L$\tablenotemark{a}} & \colhead{$V_{f=0.45}$} & \colhead{Instrument} & 
\colhead{$z$ Reference}} 
\startdata
970228 & 0.695 & $5.0\pm1.7\times10^{51}$ & $0.08\pm0.05$ & {\it Wind}/KONUS & 
Djorgovski et al. 1999b \nl
970508 & 0.835 & $1.0\pm0.2\times10^{51}$ & $0.05\pm0.02$ & {\it CGRO}/BATSE & 
Metzger et al. 1997 \nl 
970828 & 0.958 & $7.1\pm1.4\times10^{51}$ & $0.10\pm0.00$ & {\it CGRO}/BATSE & 
Djorgovski 1999 \nl 
971214 & 3.418 & $5.8\pm1.4\times10^{52}$ & $0.11\pm0.01$ & {\it CGRO}/BATSE & 
Kulkarni et al. 1998a \nl
980308 & $<4.3$\tablenotemark{b} & $<7.4\times10^{52}$ & $0.24\pm0.02$ & {\it 
CGRO}/BATSE & Schaefer et al. 1999 \nl
980326 & $<4.3$\tablenotemark{b} & $<2.5\times10^{53}$ & $0.09\pm0.02$ & {\it 
CGRO}/BATSE & Groot et al. 1998 \nl
980329 & $<3.9$\tablenotemark{c} & $<5.7\times10^{53}$ & $0.04\pm0.01$ & {\it 
CGRO}/BATSE & Djorgovski 1999 \nl
980425 & 0.0085 & $5.2\pm2.0\times10^{46}$ & $0.00\pm0.01$ & {\it CGRO}/BATSE & 
Tinney et al. 1998 \nl
980519 & $<4.3$\tablenotemark{b} & $<2.2\times10^{53}$ & $0.16\pm0.03$ & {\it 
CGRO}/BATSE & Halpern et al. 1999 \nl
980613 & 1.096 & $1.3\pm0.2\times10^{51}$ & $>0.03$\tablenotemark{d} & {\it 
CGRO}/BATSE & Djorgovski et al. 1999a \nl
980703 & 0.967 & $3.6\pm0.7\times10^{51}$ & $0.06\pm0.01$ & {\it CGRO}/BATSE & 
Djorgovski et al. 1998 \nl
981220 & $<4.3$\tablenotemark{b} & $<4.3\times10^{53}$ & $0.08\pm0.03$ & {\it 
Wind}/KONUS & Masetti et al. 1998 \nl
990123 & 1.600 & $8.5\pm1.0\times10^{52}$ & $0.10\pm0.00$ & {\it CGRO}/BATSE & 
Kulkarni et al. 1999 \nl
990510 & 1.619 & $4.3\pm0.5\times10^{52}$ & $0.24\pm0.01$ & {\it CGRO}/BATSE & 
Beuermann et al. 1999 \nl
990705 & $<5.5$\tablenotemark{b} & $<3.5\times10^{53}$ & $0.15\pm0.05$ & {\it 
Ulysses}/GRB & Masetti et al. 2000 \nl
990712 & 0.430 & $3.8\pm1.5\times10^{50}$ & $0.02\pm0.03$ & {\it Wind}/KONUS & 
Galama et al. 1999 \nl
991208 & 0.706 & $2.9\pm1.0\times10^{52}$ & $0.08\pm0.03$ & {\it Wind}/KONUS & 
Djorgovski et al. 1999c \nl
991216 & 1.020 & $1.1\pm0.2\times10^{53}$ & $0.14\pm0.01$ & {\it CGRO}/BATSE & 
Vreeswijk et al. 1999 \nl 
000131 & $<5.5$\tablenotemark{b} & $<6.7\times10^{53}$ & $0.11\pm0.01$ & {\it 
Wind}/KONUS & Pedersen et al. 2000 \nl
000301 & 2.034 & $6.2\pm2.0\times10^{52}$ & $>0.03$\tablenotemark{d} & 
NEAR/XGRS\tablenotemark{e} & Castro et al. 2000 \nl
\enddata
\tablenotetext{a}{Isotropic-equivalent peak photon luminosity in erg s$^{-1}$
between source-frame energies 100 and 1000 keV, for peak fluxes measured
on a 1-s timescale, $H_0 = 65$ km s$^{-1}$ Mpc$^{-1}$, $\Omega_m =
0.3$, and $\Omega_{\Lambda} = 0.7$; upper limits are 1 $\sigma$.}
\tablenotetext{b}{From non-detection of Lyman limit in afterglow or host galaxy 
photometry.}
\tablenotetext{c}{From non-detection of Ly$\alpha$ forest in host galaxy 
spectroscopy.}
\tablenotetext{d}{Only 1-s resolution light curve is available; lower limit is 1 
$\sigma$.}
\tablenotetext{e}{Peak flux from {\it Ulysses}/GRB.}
\end{deluxetable}

\clearpage

\begin{deluxetable}{cccccc}
\footnotesize
\tablecolumns{5}
\tablewidth{0pc}
\tablecaption{Smoothing Timescales}
\tablehead{\colhead{GRB} & \colhead{Instrument\tablenotemark{a}} & 
\colhead{$T_{f=0.45}^{E_1}$\tablenotemark{b}} & 
\colhead{$T_{f=0.45}^{E_2}$\tablenotemark{b}} &
\colhead{$T_{f=0.45}^{E_3}$\tablenotemark{b}} & 
\colhead{$T_{f=0.45}^{E_4}$\tablenotemark{b}}} 
\startdata
970228 & {\it Wind}/KONUS & 3.235 & 2.891 & 1.061 & --\tablenotemark{c} \nl
970508 & {\it CGRO}/BATSE & 3.711 & 3.114 & 2.944 & 1.333 \nl 
970828 & {\it CGRO}/BATSE & 19.057 & 15.875 & 12.335 & 11.276 \nl 
971214 & {\it CGRO}/BATSE & 7.769 & 6.706 & 5.943 & 2.080 \nl
980308 & {\it CGRO}/BATSE & 4.987 & 6.135 & 4.903 & --\tablenotemark{d} \nl
980326 & {\it CGRO}/BATSE & 0.782 & 0.836 & --\tablenotemark{e} & 
--\tablenotemark{d} \nl
980329 & {\it CGRO}/BATSE & 5.371 & 5.033 & 4.769 & 4.382 \nl
980425 & {\it CGRO}/BATSE & 7.819 & 6.701 & 5.169 & 0.839 \nl
980519 & {\it CGRO}/BATSE & 8.286 & 7.232 & 4.742 & 2.872 \nl
980613 & {\it CGRO}/BATSE & 8.160 & 8.432 & 6.528 & --\tablenotemark{d} \nl
980703 & {\it CGRO}/BATSE & 18.024 & 16.659 & 15.191 & 11.733 \nl
981220 & {\it Wind}/KONUS & 2.152 & 1.871 & 0.740 & --\tablenotemark{c} \nl
990123 & {\it CGRO}/BATSE & 22.673 & 19.346 & 13.340 & 7.345 \nl
990510 & {\it CGRO}/BATSE & 6.911 & 4.964 & 3.118 & --\tablenotemark{d}  \nl
990705 & {\it Ulysses}/GRB & 4.986 & --\tablenotemark{f} & --\tablenotemark{f} & 
--\tablenotemark{f} \nl
990712 & {\it Wind}/KONUS & 2.405 & 3.595 & 0.925 & --\tablenotemark{c} \nl
991208 & {\it Wind}/KONUS & 7.737 & 4.682 & 4.299 & --\tablenotemark{c} \nl
991216 & {\it CGRO}/BATSE & 4.641 & 3.842 & 2.905 & 2.009 \nl 
000131 & {\it Wind}/KONUS & 6.151 & 4.632 & 2.810 & --\tablenotemark{c}  \nl
000301 & NEAR/XGRS & 1.552 & --\tablenotemark{f} & --\tablenotemark{f} & 
--\tablenotemark{f} \nl 
\enddata
\tablenotetext{a}{{\it CGRO}/BATSE:  25 keV $\la E_1 \la$ 55 keV, 55 keV $\la 
E_2 \la$ 110 keV, 110 keV $\la E_3 \la$ 320 keV, $E_4 \ga 320$ keV; {\it 
Wind}/KONUS:  10 keV $\la E_1 \la$ 45 keV, 45 keV $\la E_2 \la$ 190 keV, 190 keV 
$\la E_3 \la$ 770 keV; {\it Ulysses}/GRB:  22 keV $\la E_1 \la$ 150 keV; 
NEAR/XGRS:  300 keV $\la E_1 \la$ 1000 keV.}
\tablenotetext{b}{45\% smoothing timescale in seconds of light curve in 
observer-frame energy band $E_i$.}
\tablenotetext{c}{Only three energy bands are available for this instrument.}
\tablenotetext{d}{The smoothing timescale of the light curve in this energy band 
is less than the imposed effective sampling timescale of the light curve due to 
negligible emission in this energy band; consequently, we drop this energy 
band.}
\tablenotetext{e}{There is an unusual, positive, constant, systematic offset of 
a portion of the light curve in this energy band; consequently, we drop this 
energy band.}
\tablenotetext{f}{Only one energy band is available for this instrument.}
\end{deluxetable}

\clearpage

\figcaption[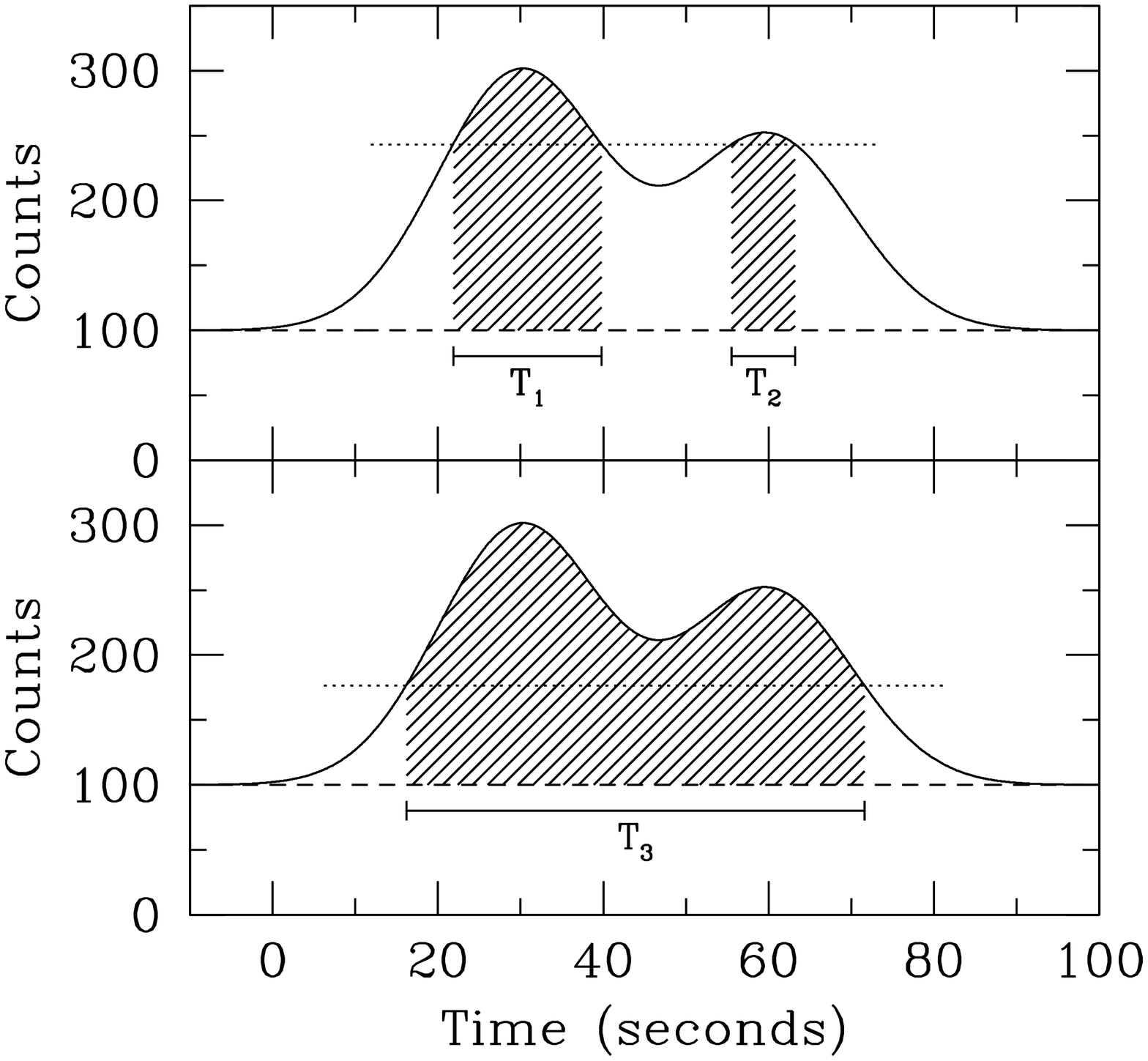]{A schematic illustration of our measure of
burst duration.  The solid line represents a burst light curve,
acquired in observer-frame energy band $E$, and the dashed line
represents the background.  Top panel:  The hashed area is the
brightest 50\% of the total counts above background.  Hence, the 50\%
duration, or smoothing timescale, of this light curve is given by
$T_{f=0.5}^E = T_1 + T_2$.  Bottom panel:  The hashed area is the
brightest 90\% of the total counts above background.  Hence, the 90\%
duration, or smoothing timescale, of this light curve is given by
$T_{f=0.9}^E = T_3$.\label{cepheid1.eps}}

\figcaption[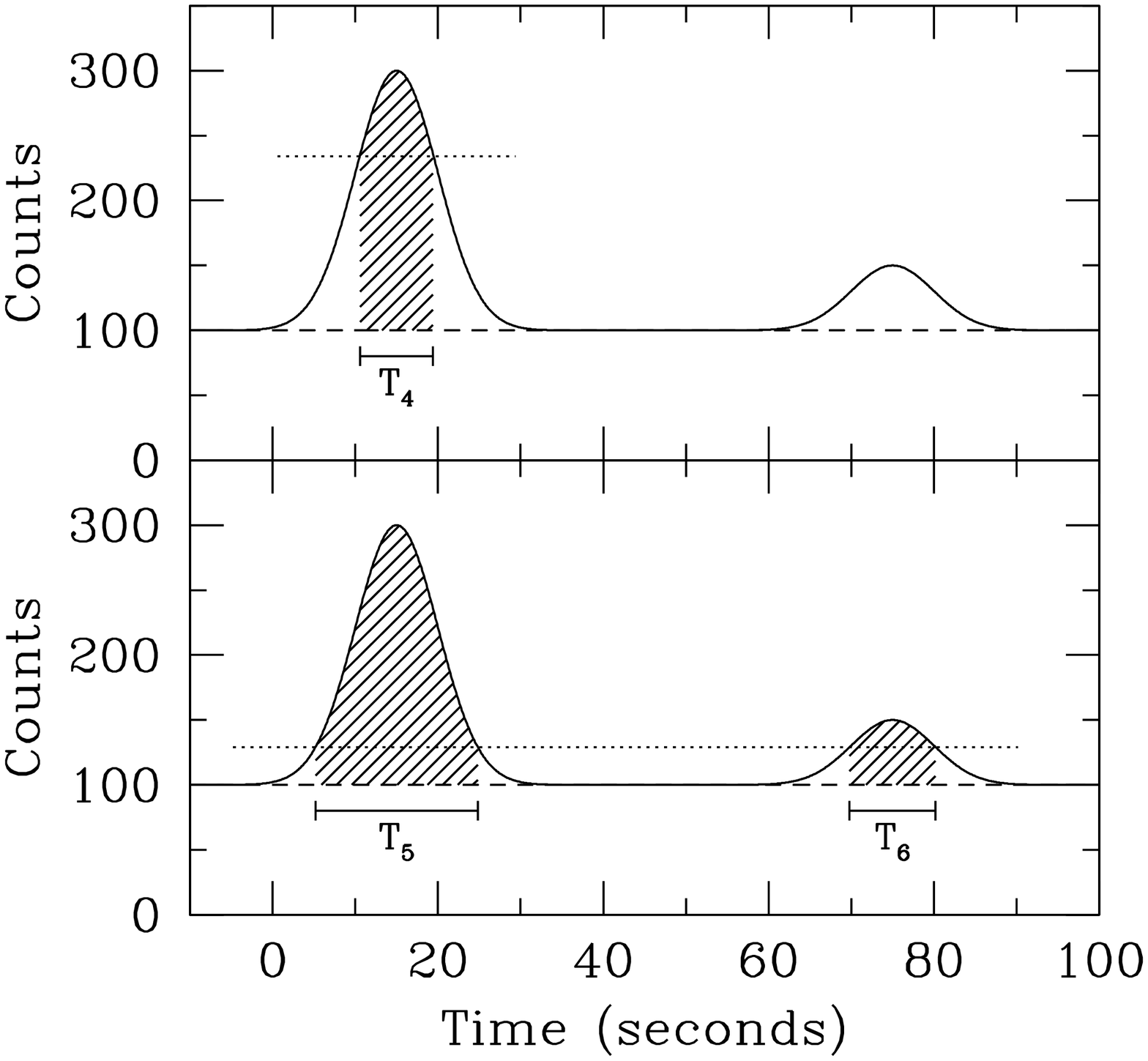]{Another schematic illustration of our measure
of burst duration.  The solid line represents a burst light curve,
acquired in observer-frame energy band $E$, and the dashed line
represents the background.  Top panel:  The hashed area is the
brightest 50\% of the total counts above background.  Hence, the 50\%
duration, or smoothing timescale, of this light curve is given by
$T_{f=0.5}^E = T_4$.  Bottom panel:  The hashed area is the brightest
90\% of the total counts above background.  Hence, the 90\% duration,
or smoothing timescale, of this light curve is given by $T_{f=0.9}^E =
T_5 + T_6$.\label{cepheid2.eps}}

\figcaption[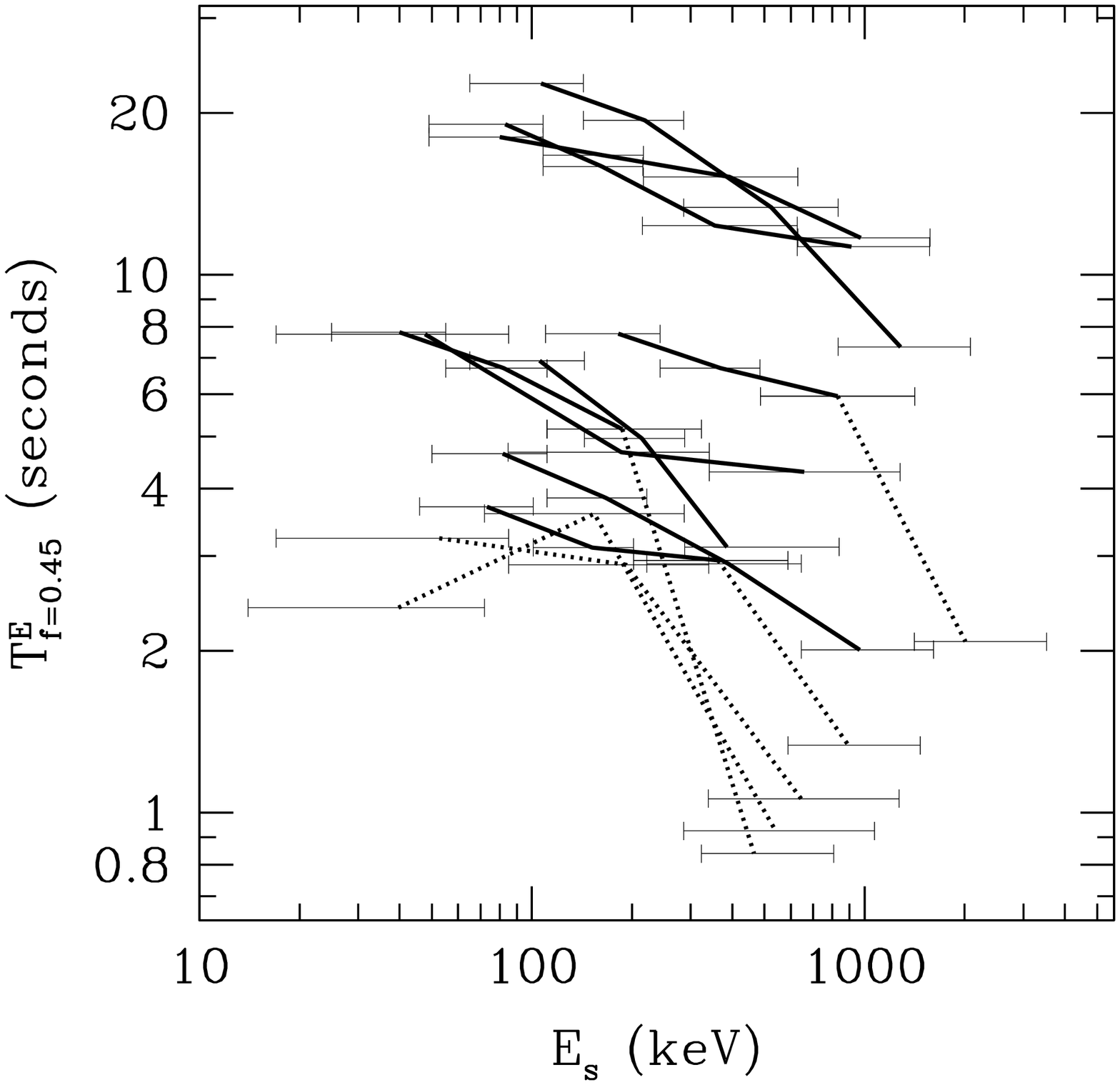]{Bursts in our sample for which spectroscopic
redshifts, peak fluxes, and 64-ms or better resolution light curves are
available.  For each of these bursts, light curves were acquired in
typically 3 or 4 independent energy bands $E$.  For each burst, we
compute the durations, or smoothing timescales, $T_{f=0.45}^E$ of its 3
or 4 independent light curves, and plot $T_{f=0.45}^E$ as a function of
source-frame energy band $E_s = E(1+z)$.  The horizontal bars mark the
source-frame energy bands.  We plot solid lines if the light curve has
$> 3000$ total counts above background, and dotted lines if the light
curve has $< 3000$ total counts above background.  The dotted portions
of these curves suffer from low signal-to-noise; very little
variability information can be gleaned from these light curves. 
However, the trend is clear:  the durations, or smoothing timescales,
$T_{f=0.45}^E$ of these bursts are shorter at higher energies.  From
the solid portions of these curves, we find that $T_{f=0.45}^E \sim
E^{-0.4}$, in agreement with the principle result of Fenimore et al.
(1995).  Other values of $f$ yield similar
results.\label{cepheid3.eps}}

\figcaption[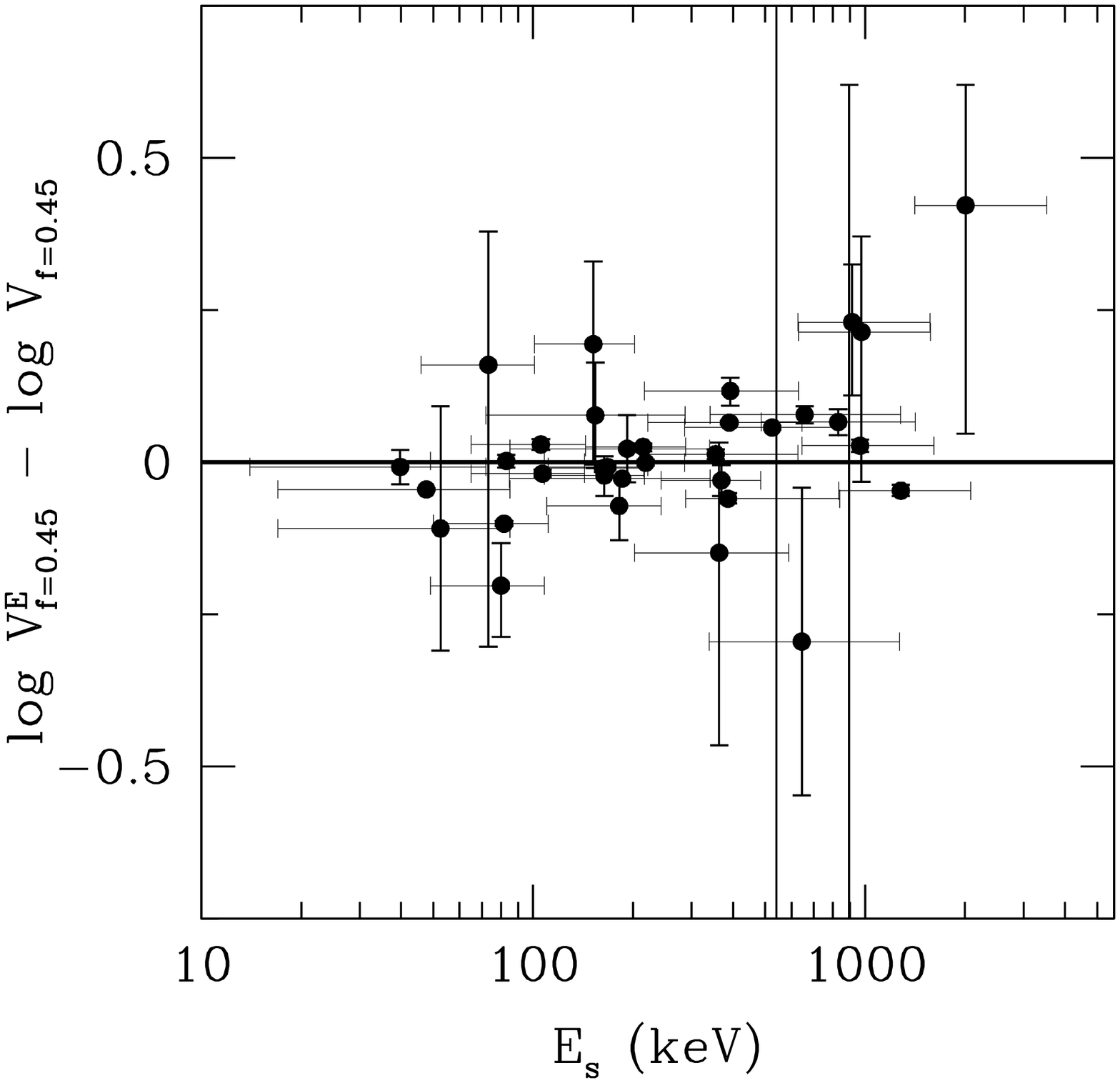]{Bursts in our sample for which spectroscopic
redshifts, peak fluxes, and 64-ms or better resolution light curves are
available.  For each of these bursts, light curves were acquired in
typically 3 or 4 independent energy bands $E$.  For each burst, we
compute the variabilities $V_{f=0.45}^E$ of its 3 or 4 independent
light curves, and the combined variability $V_{f=0.45}$ of the burst. 
We plot $\log{V_{f=0.45}^E} - \log{V_{f=0.45}}$ versus source-frame
energy band $E_s = E(1+z)$.  The horizontal bars mark the source-frame
energy bands.  Clearly, there is no significant trend with energy. 
Other values of $f$ yield similar results.\label{cepheid4.eps}}

\figcaption[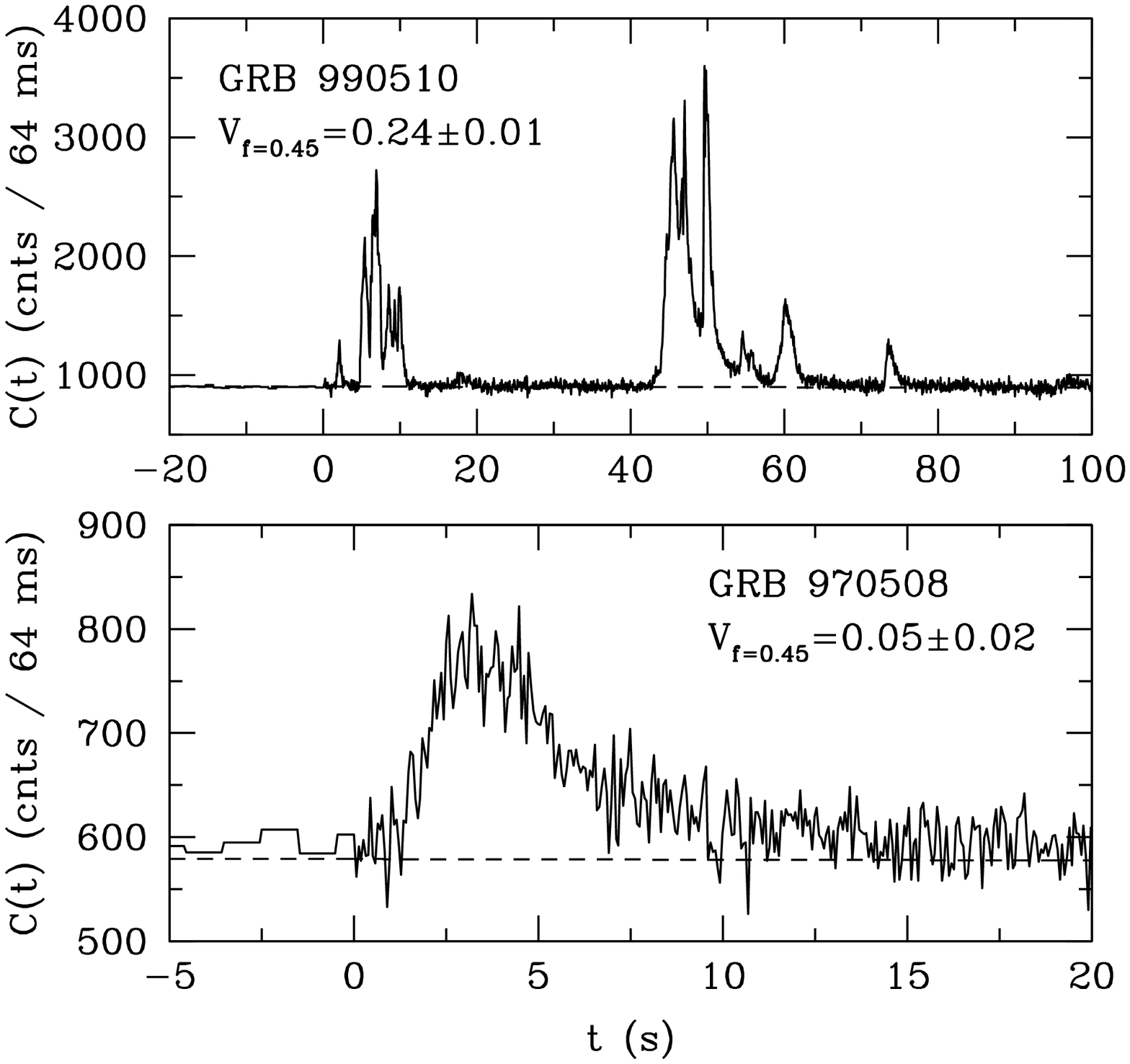]{The $> 25$ keV light curves of the most (GRB 990510) 
and least (GRB 970508) variable cosmological BATSE bursts in our sample.  In the 
case of GRB 990510 ($z = 1.619$), we find that $V_{f=0.45} = 0.24 \pm 0.01$.  In 
the case of GRB 970508 ($z = 0.835$), we find that $V_{f=0.45} = 0.05 \pm 
0.02$.\label{cepheid4b.eps}}

\figcaption[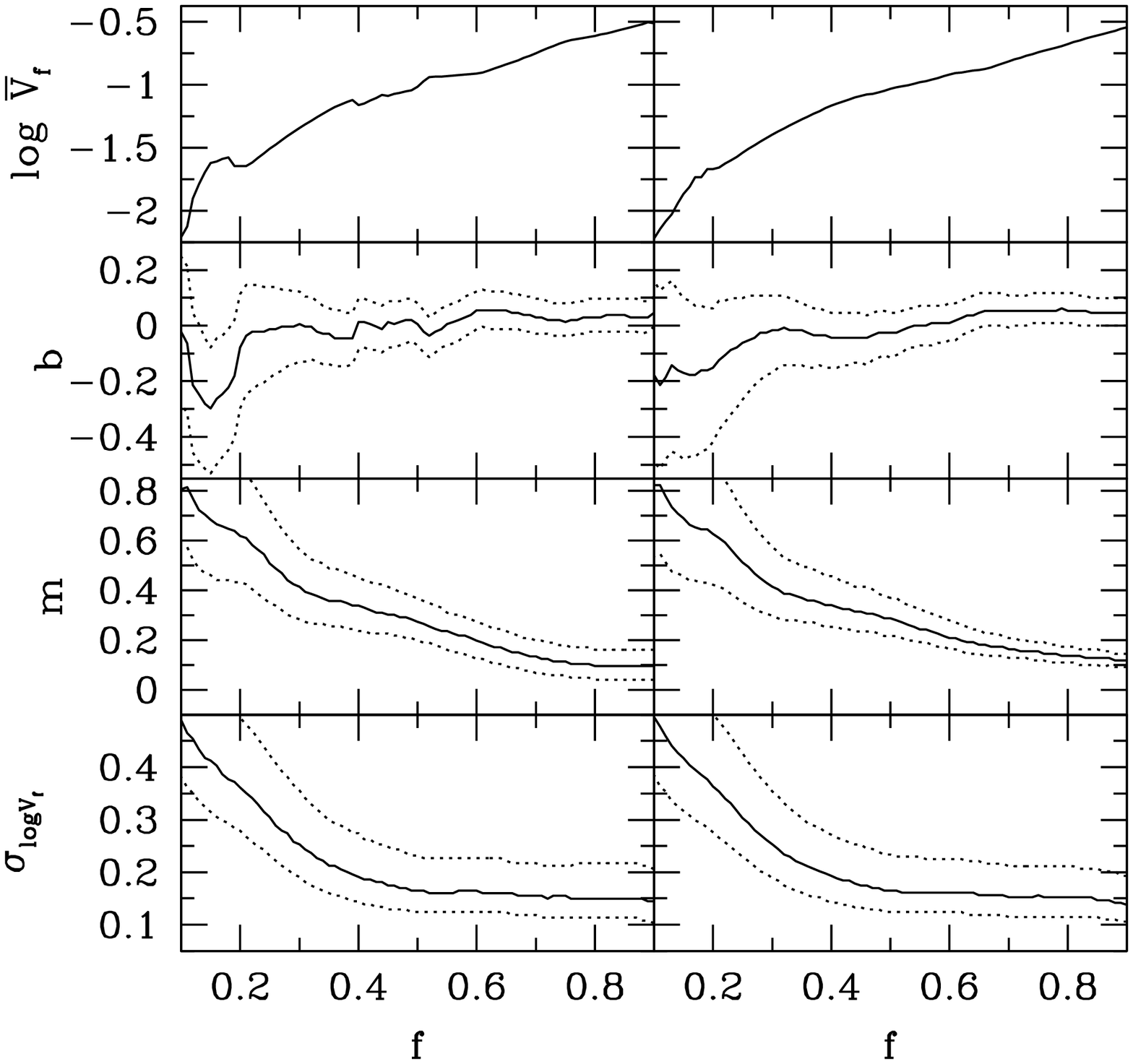]{The value of $\log{{\bar{V}}_f}$ as a function of $f$, 
and the best-fit values, and 1 $\sigma$ uncertainties in the best-fit values, of 
the model parameters $b$, $m$, and $\sigma_{\log{V_f}}$ as functions of $f$.  In 
the left panels, we exclude GRB 980425 from the fits; in the right panels we 
include GRB 980425 in the fits.\label{cepheid5.eps}}

\figcaption[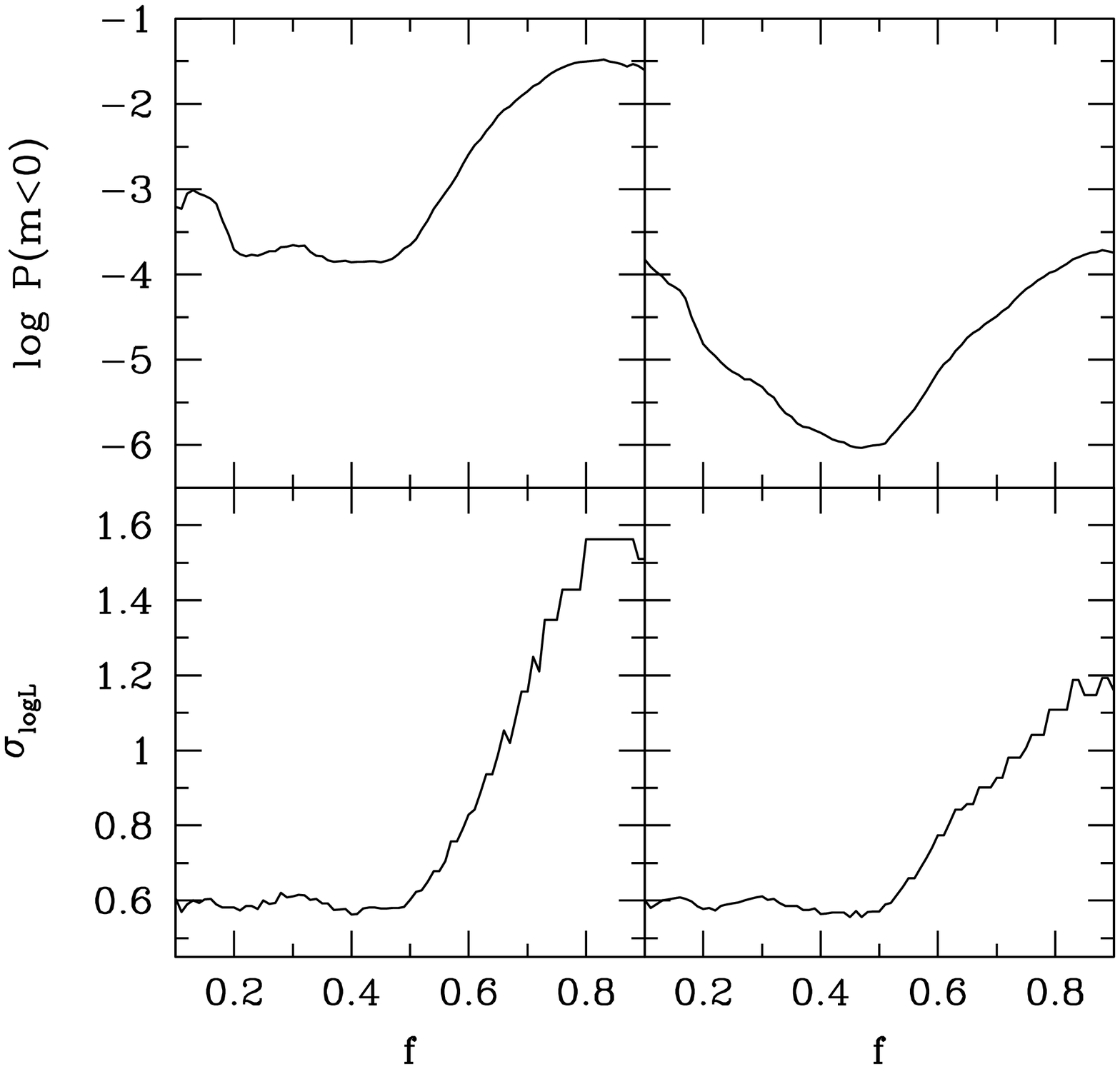]{The probability $P(m<0)$ that $m < 0$ as a function of 
$f$, and the extrinsic scatter $\sigma_{\log{L}}$ of the data along the 
$\log{L}$ axis as a function of $f$.  In the left panels, we exclude GRB 980425 
from the fits; in the right panels we include GRB 980425 in the 
fits.\label{cepheid6.eps}}

\figcaption[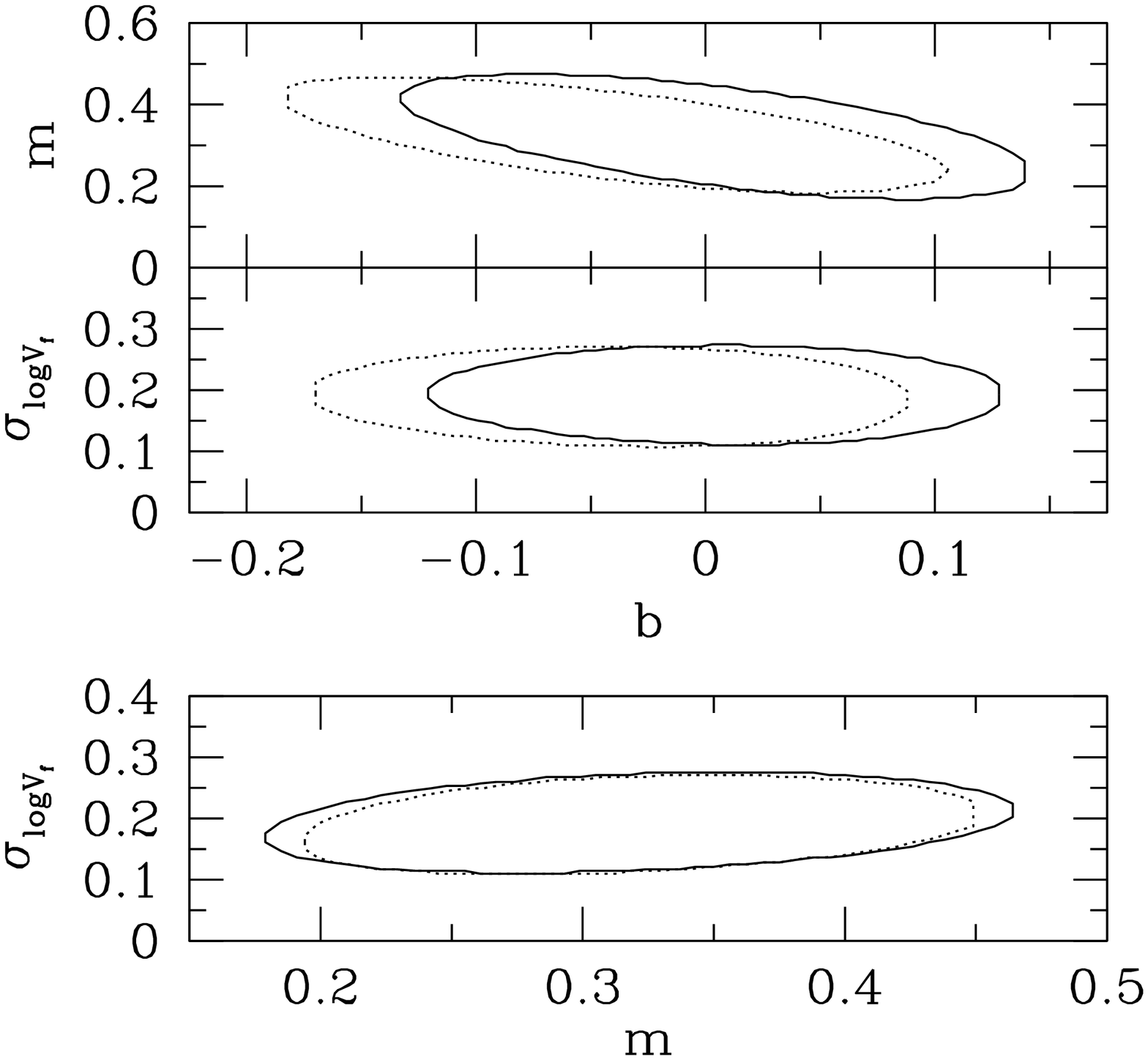]{The two-dimensional uncertainty distributions
of the fitted values of the model parameters $b$, $m$, and
$\sigma_{\log{V_f}}$.  The curves mark the 1 $\sigma$ credible
regions.  The solid curves are for the case in which we exclude
GRB 980425 from the fits ($f = 0.45$); the dotted curves are for the case in 
which we include GRB 980425 in the fits ($f = 0.47$).\label{cepheid7.eps}}

\figcaption[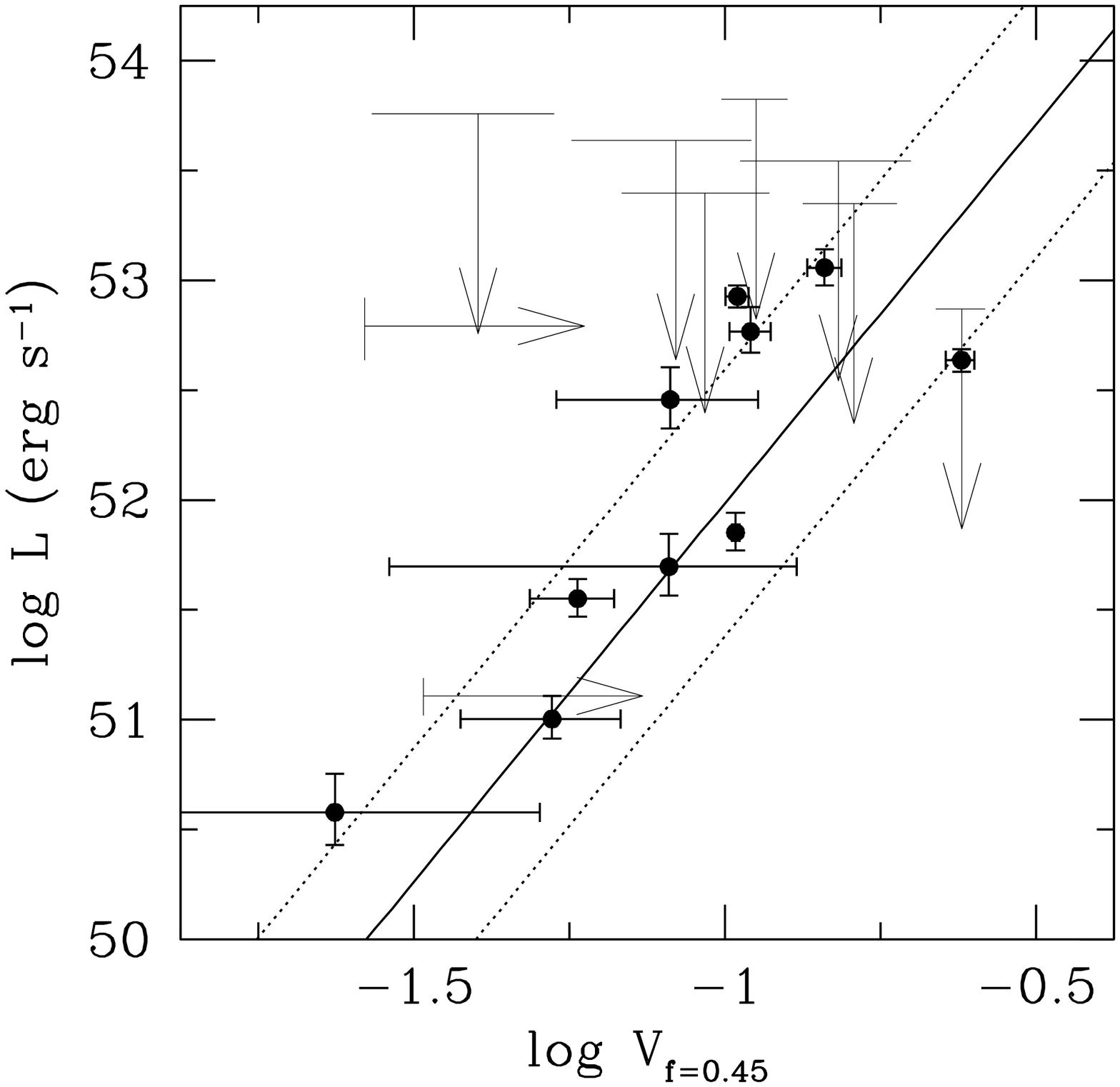]{The variabilities $V_{f=0.45}$ and 
isotropic-equivalent peak luminosities $L$ of the bursts in our sample, 
excluding GRB 980425.  The solid and dotted lines mark the center and 1 $\sigma$ 
widths of the best-fit model distribution of these bursts in the 
$\log{L}$-$\log{V_{f=0.45}}$ plane.\label{cepheid8.eps}}

\figcaption[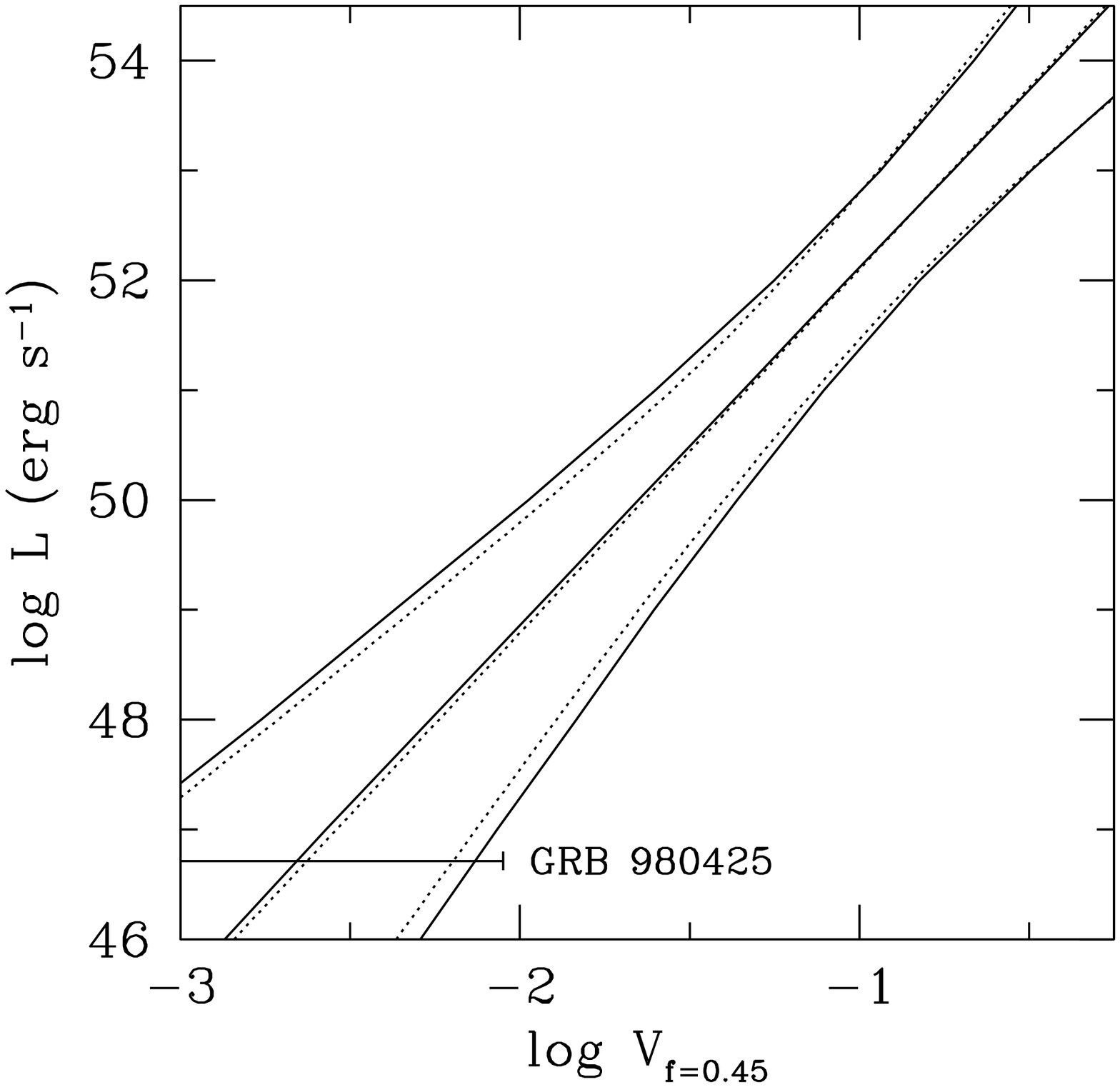]{The solid lines mark the best estimate for $L$ as a 
function of $V_{f=0.45}$, and the 1 $\sigma$ uncertainty in $L$ as a function of 
$V_{f=0.45}$ for the case in which we exclude GRB 980425 from the fits.  The 
dotted lines mark analytic approximations to these functions, given by Equations 
(\ref{line}) and (\ref{lumerr}) and the fitted values of the model parameters.  
Clearly, GRB 980425 is consistent with the fitted model.\label{cepheid9.eps}}

\figcaption[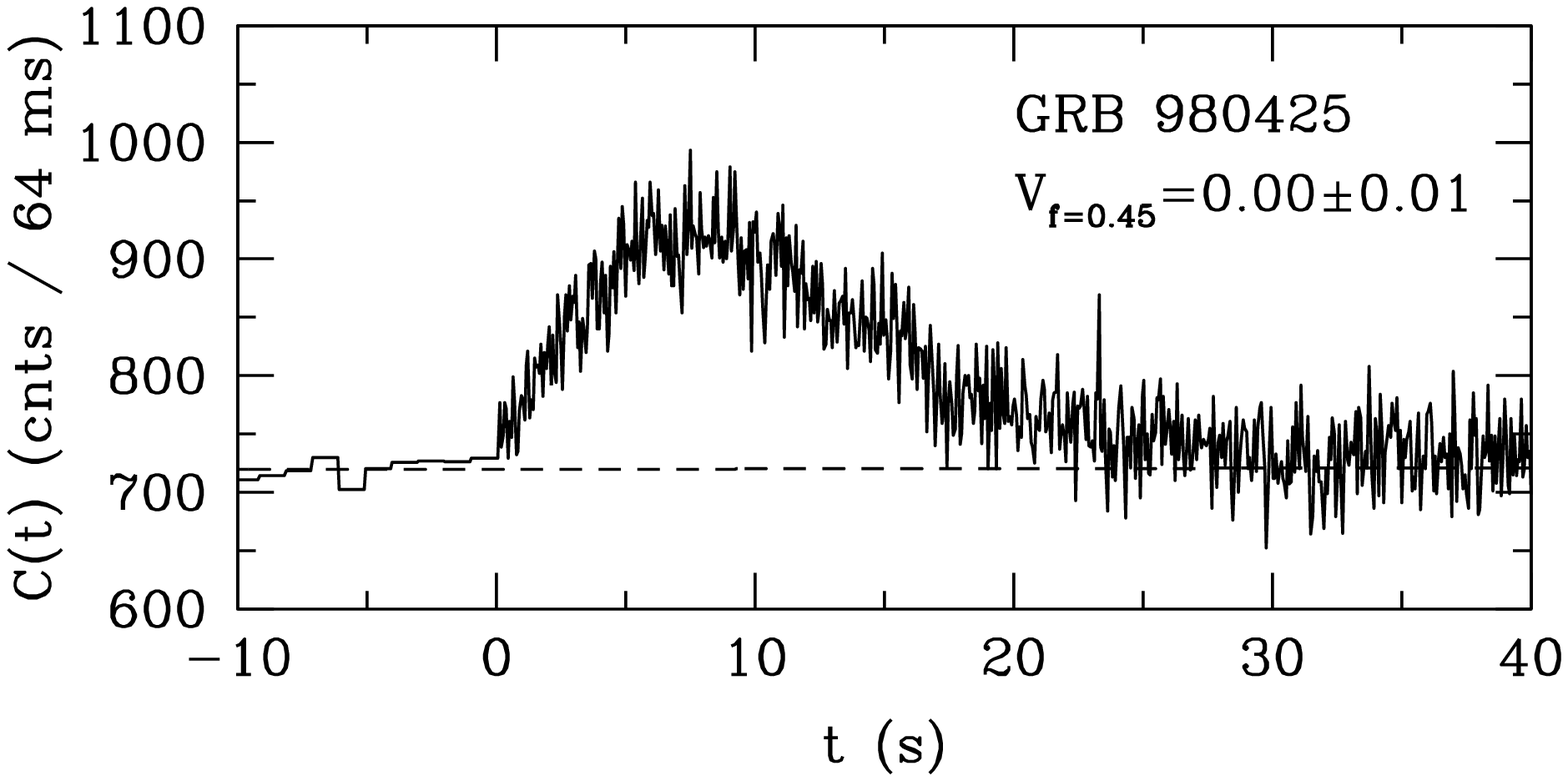]{The $> 25$ keV BATSE light curve of GRB 980425.  We 
find that $V_{f=0.45} = 0.00 \pm 0.01$. \label{cepheid9b.eps}}

\figcaption[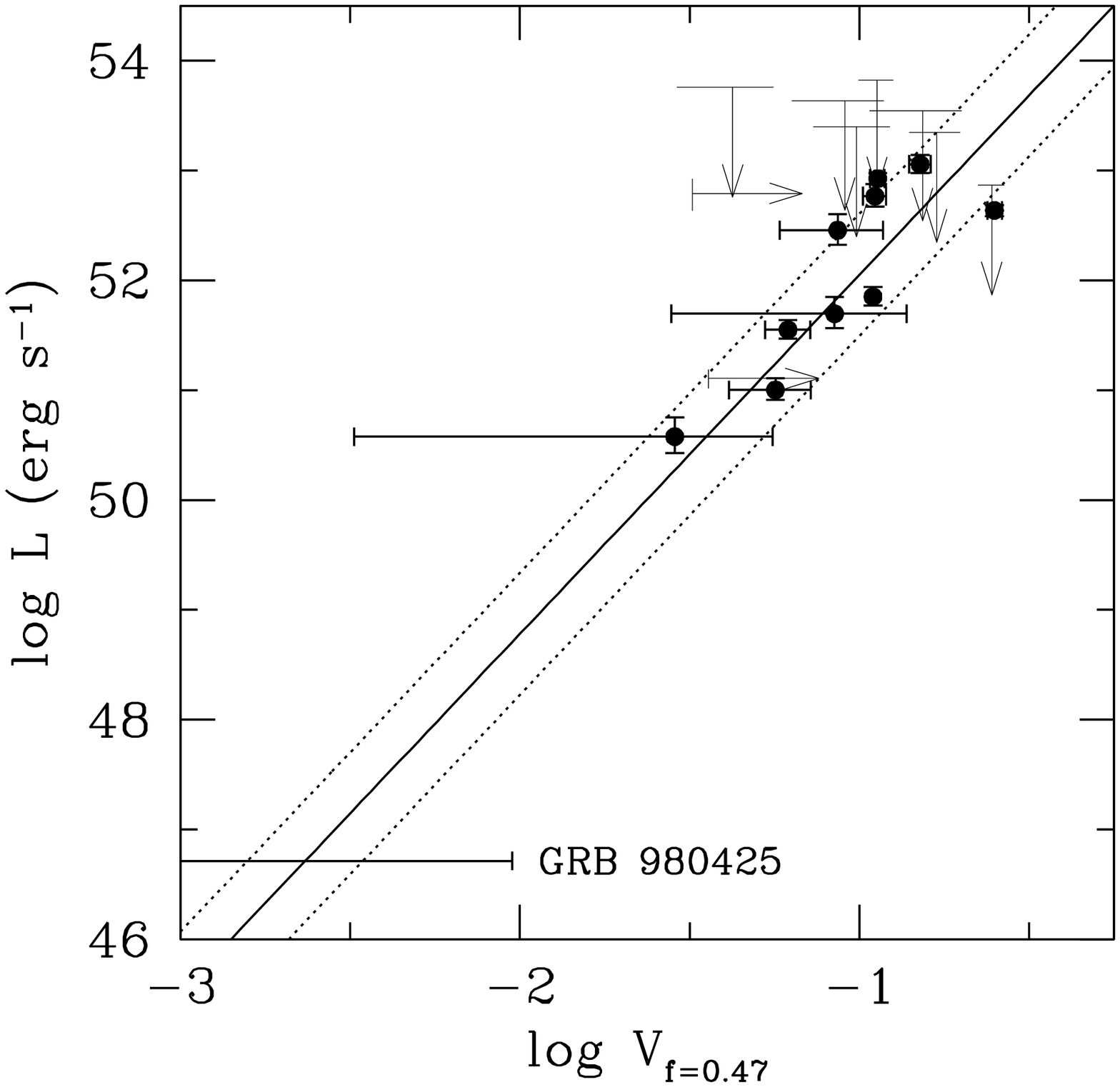]{The variabilities $V_{f=0.47}$ and 
isotropic-equivalent peak luminosities $L$ of the bursts in our sample, 
including GRB 980425.  The solid and dotted lines mark the center and 1 $\sigma$ 
widths of the best-fit model distribution of these bursts in the 
$\log{L}$-$\log{V_{f=0.47}}$ plane.\label{cepheid10.eps}}

\figcaption[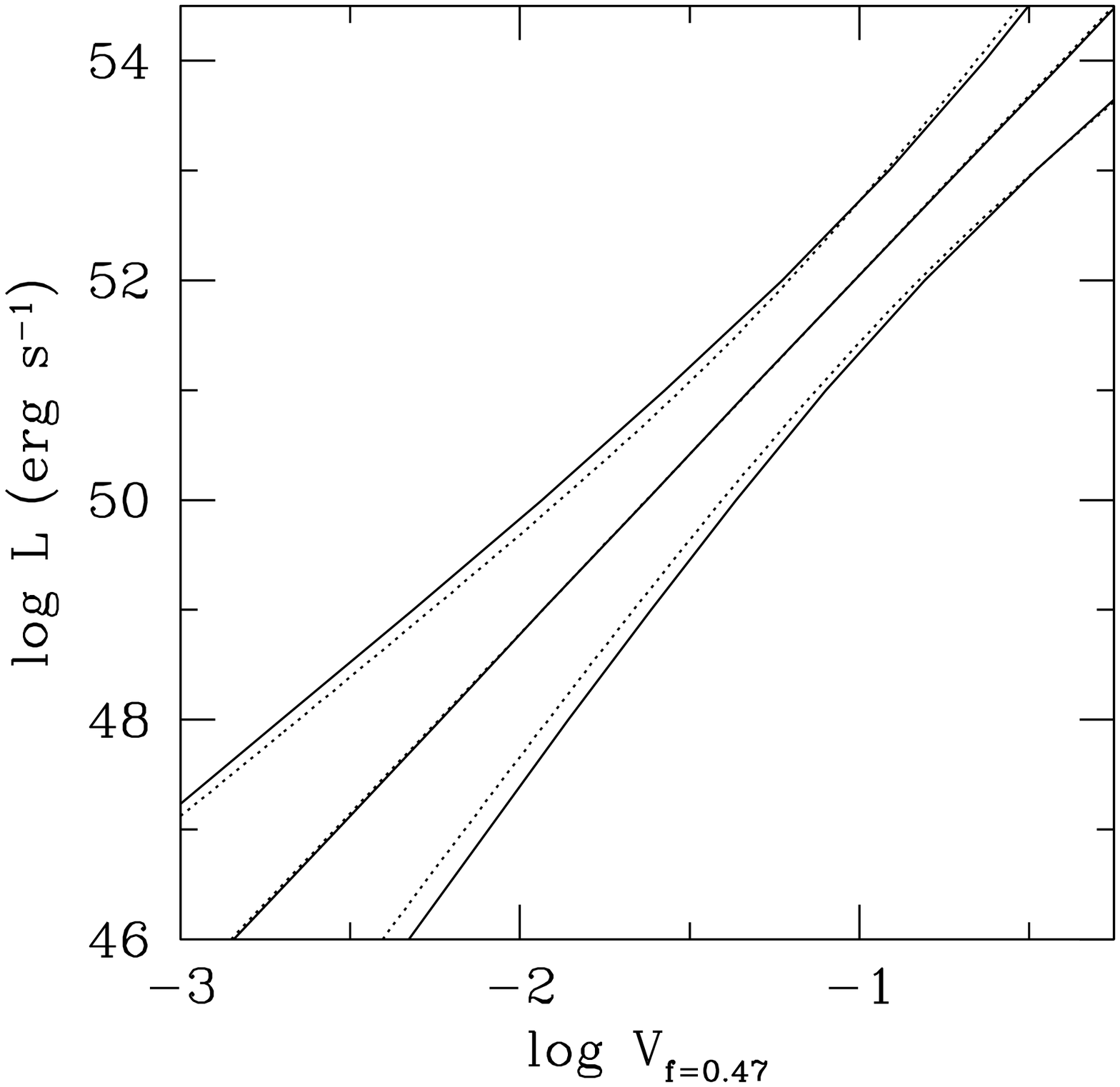]{The solid lines mark the best estimate for
$L$ as a function of $V_{f=0.47}$, and the 1 $\sigma$ uncertainty in
$L$ as a function of $V_{f=0.47}$ for the case in which we include GRB
980425 in the fits.  The dotted lines mark analytic approximations to
these functions, given by Equations (\ref{line}) and
(\ref{lumerr}) and the fitted values of the model 
parameters.\label{cepheid11.eps}}

\clearpage

\setcounter{figure}{0}

\begin{figure}[tb]
\plotone{cepheid1.eps}
\end{figure}

\begin{figure}[tb]
\plotone{cepheid2.eps}
\end{figure}

\begin{figure}[tb]
\plotone{cepheid3.eps}
\end{figure}

\begin{figure}[tb]
\plotone{cepheid4.eps}
\end{figure}

\begin{figure}[tb]
\plotone{cepheid4b.eps}
\end{figure}

\begin{figure}[tb]
\plotone{cepheid5.eps}
\end{figure}

\begin{figure}[tb]
\plotone{cepheid6.eps}
\end{figure}

\begin{figure}[tb]
\plotone{cepheid7.eps}
\end{figure}

\begin{figure}[tb]
\plotone{cepheid8.eps}
\end{figure}

\begin{figure}[tb]
\plotone{cepheid9.eps}
\end{figure}

\begin{figure}[tb]
\plotone{cepheid9b.eps}
\end{figure}

\begin{figure}[tb]
\plotone{cepheid10.eps}
\end{figure}

\begin{figure}[tb]
\plotone{cepheid11.eps}
\end{figure}
\end{document}